\documentclass[fleqn,twoside]{article}
\usepackage{amssymb}
\usepackage{macros}
\usepackage{graphicx}
\usepackage{espcrc2}

\title{\begin{flushleft} \normalsize{DESY 01-150\hfill hep-lat/0110060} \end{flushleft}The $\nf=0$ heavy quark potential and perturbation theory}
\author{Silvia Necco\address{DESY, Platanenallee 6, D-15738 Zeuthen, Germany}\thanks{This work was supported by the European Community's Human potential programme under HPRN-CT-2000-00145 Hadrons/LatticeQCD.}}
\begin{document}
\begin{abstract}
The potential of a static quark-antiquark pair is studied in the range 
$0.05$$\,\fm\leq r\leq 0.8$$\,\fm$,
employing a sequence of lattices up to $64^4$. The continuum quantities are 
evaluated by extrapolation of the data at finite lattice spacing. The results are compared with the perturbation theory predictions obtained from the solution of the renormalization group equation for the coupling. The renormalization scheme must be chosen carefully. No evidence for large 
non-perturbative effects at short distances is seen.
\end{abstract}
\maketitle
\vspace{-1cm}
\section{THE RUNNING COUPLING IN PERTURBATION THEORY}
The potential between static quarks in the fundamental representation defines a physical renormalized coupling, $\alpha_{V}(\mu)$, through
\be
\tilde{V}(Q)=-4\pi C_{F}\frac{\alpha_{\rm {V}}(\mu)}{Q^2}\, , \, \mu=Q,
\ee
in the momentum space, or alternatively
\be\label{pot_coord}
V(r)=-C_{F}\frac{\alpha_{\overline{\rm V}}(\mu)}{r}\, , \, \mu=1/r,
\ee
from which one defines the force
\be\label{force}
F(r)=\frac{\rm{d}V}{\rm{d}r}=C_{F}\frac{\alpha_{\rm {qq}}(\mu)}{r^2}\, ,\,\mu=1/r.
\ee
A perturbative prediction for the running coupling in a given renormalization scheme can be obtained through the $\Lambda$-parameter
\be\label{lambda}
\lambdas=\mu(b_{0}\bar{g}^2)^{-b_{1}/2b_{0}^2}e^{-1/2 b_{0}\bar{g}^2}\cdot
\ee
$$
\exp
\left\{-\int_{0}^{\bar{g}}dx\left[\frac{1}{\beta(x)}+\frac{1}{b_{0}x^3}-\frac{b_{1}}{b_{0}^{2}x}\right]\right\},
$$
where $\bar{g}=(4\pi\alpha(\mu))^{1/2}$, and
\be
\beta(\bar{g})=\mu\frac{\rm{d}\bar{g}}{\rm{d}\mu}
\ee
$$
\sim_{\bar{g}\rightarrow 0}\quad\bar{g}^3\left\{b_{0}+b_{1}\bar{g}^2+b_{2}\bar{g}^4+...\right\}
$$
\be
b_0=\frac{11}{16\pi^2},
              \, b_1=\frac{51}{128\pi^4}, \,(\nf=0).
\ee
Due to the fact that the $\Lambda$- parameter,
\be
\lambdaMSbar\rnod=0.602(48),\,(\nf=0)
\ee
is known \cite{lambdams}
with $\rnod\approx 0.5\fm$, and that the relation between $\Lambda$ parameters in different schemes is exactly determined by one-loop coefficients, \eq{lambda} furnishes a parameter-free prediction for $\alpha$. In particular, truncating $\beta$ after the term $b_{n-1}$ and solving numerically for $\bar{g}(\mu)$ defines the $n$-loop RG solution for the coupling.

\begin{figure}
\includegraphics[width=15pc]{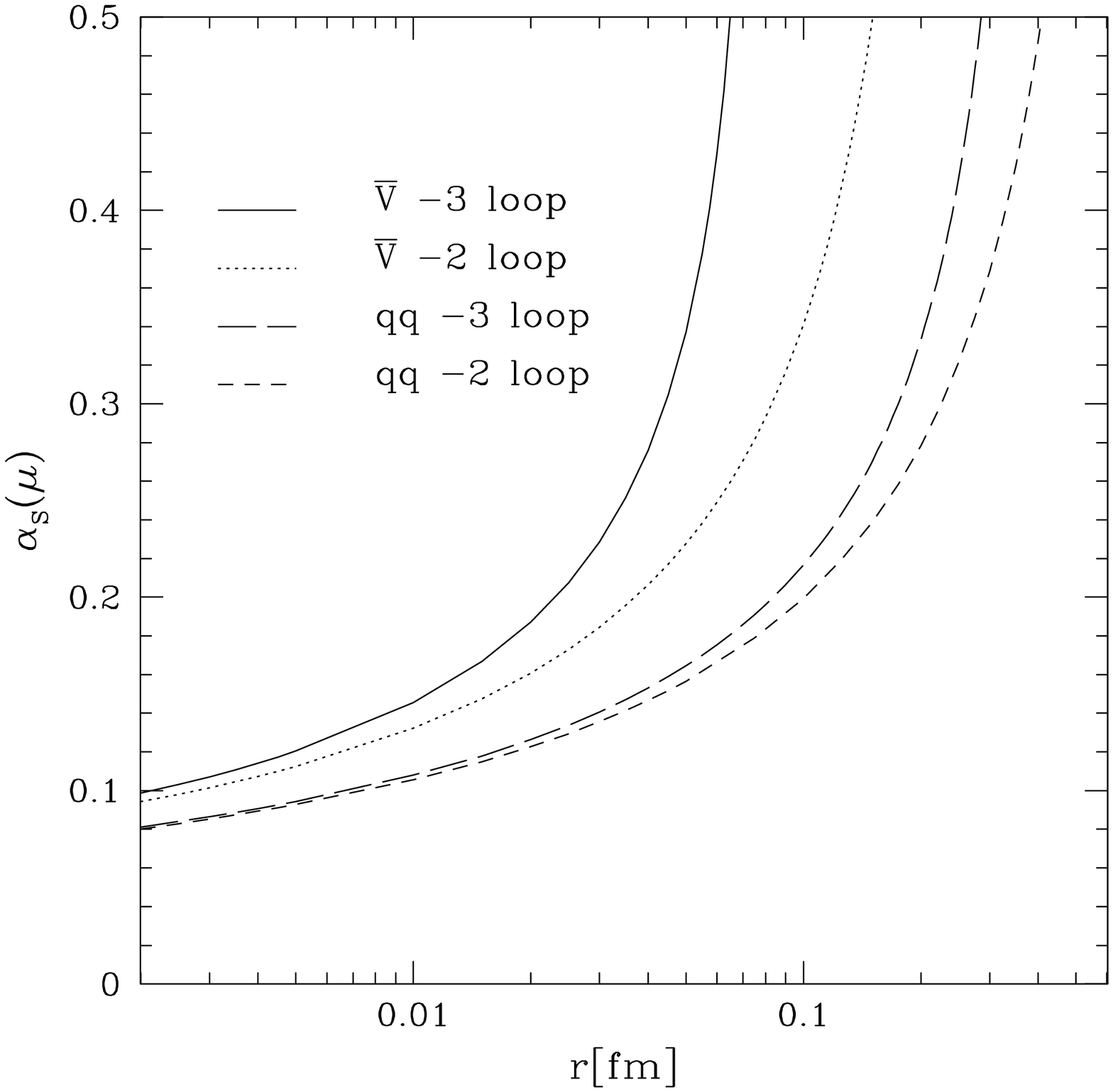}
\vspace{-1cm}
\caption{Running couplings obtained by integration of the RG at 2- and
3-loops with $\lambdaMSbar=238\MeV$.}\label{pertu}
\end{figure}

The ratios of $\Lambda$ parameters and the 3-loop coefficients of the $\beta$-function for different schemes are listed in Table~\ref{coeff}. In general, large values of $b_{2}$ indicate a bad choice of renormalization scheme. As Fig~\ref{pertu} shows, evaluating $\alpha_{\rm{qq}}$ from 2- and 3-loop RG one observes that the perturbative expansion appears quite well behaved up to distances $r\approx 0.25\fm$. This corresponds to $\alpha\approx 0.25$, where the difference between 3- and 2- loops is about $10\%$. A similar analysis for $\alpha_{\overline{\rm V}}$ yields that in this scheme the perturbative expansion seems to be applicabile only up to $\alpha\approx 0.15$.
\vspace{-0.5cm}
\begin{table}[h]
\caption{Ratio of $\Lambda$-parameters and 3-loops coefficient of the $\beta$-function for various schemes for $\nf=0$ \cite{peter,schroder,qq,lambdasf,b2sf}.}
\label{coeff}
\begin{tabular}{l| l  l l  l }
\hline
scheme:       & $\rm {qq}$  & $\rm V$  & $\overline{\rm V}$  &  $\rm {SF}$  \\
\hline
& & & &\\
 $\Lambda_{\rm S}/\lambdaMSbar$ &  $e^{\gamma-35/66}$ & $e^{31/66}$ & $e^{31/66+\gamma}$ &$0.49$\\  
 $b_{2}^{\rm S}\cdot(4\pi)^3$   & $1.6524$ & $2.1287$ & $4.3353$ & $0.48$\\
\hline
\end{tabular}
\end{table}

\section{LATTICE SIMULATION}
\begin{figure}[h]
\includegraphics[width=15pc]{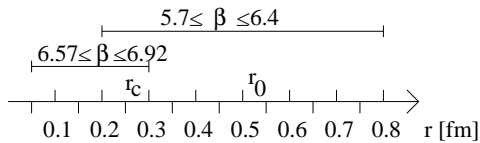}\label{rcr0}
\vspace{-0.8cm}
\caption{Two ranges of simulations: $5.7\leq\beta\leq 6.4$ \cite{oldsim} and $6.57\leq\beta\leq6.92$ \cite{silvia1}.}
\end{figure}

The physical scale is set by \cite{r0}
\be
r^{2}F(r)|_{r=r(c)}=c,
\ee
$$
\rnod=r(1.65)\approx 0.5\fm\, , \rc=r(0.65)\approx 0.26\fm.
$$
The evaluation of the static quark potential is performed over two regions with a small overlap, in order to cover the range from short to intermediate distances, as shown in Fig~\ref{rcr0}.

\vspace{-0.8cm}
This allows a continuum extrapolation for the force and the potential in the whole range $0.1\,\fm\leq r \leq 0.8\,\fm$. Data for finite $a$ can also be included, when the discretization error is estimated to be smaller than the statistical uncertainty.
The details of the lattice simulation are reported in \cite{silvia1}.

\section{COMPARISON}
\begin{figure}[t]
\includegraphics[width=15pc]{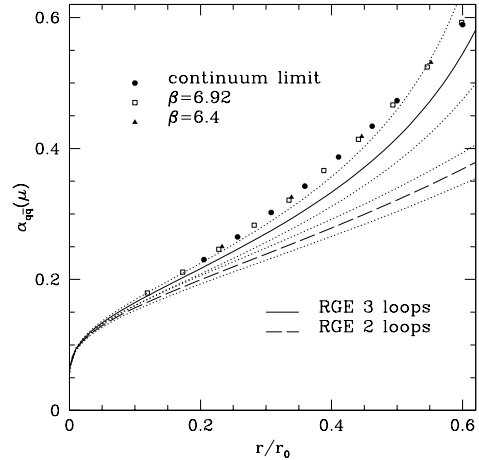}
\vspace{-1cm}
\caption{Running coupling in the $\rm{qq}$ scheme. The errors are smaller than the size of the symbols. The dotted lines correspond to the uncertainty in $\Lambda\rnod$.}\label{alpha_qq}
\end{figure}
\begin{figure}
\includegraphics[width=15pc]{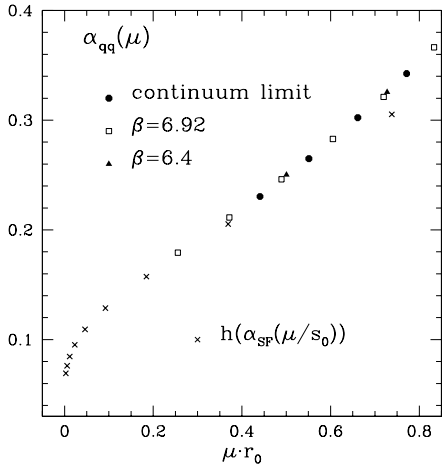}
\vspace{-1cm}
\caption{Matching with $\rm{SF}$ scheme. The non-perturbative values for $\alpha_{\rm{SF}}(\mu)$ are extracted from \cite{lambdams}.}\label{alpha_qqsf}
\end{figure}
Amongst other questions, we are interested to investigate whether there are large non perturbative terms in the short distance potential. 
We assume the following definition for ``large non-perturbative terms''.
\begin{itemize}
\item[(i)] A certain quantity is considered in a region where its perturbative expansion looks well behaved, i.e. the $n$-loop contribution is significant smaller than the $(n-1)$-loop contribution.
\item[(ii)] The difference between the full non-perturbative observable and the truncated perturbative series is much larger than the last term in the series.
\end{itemize} 
In Fig.~\ref{alpha_qq} the results for $\alpha_{\rm{qq}}$ are compared with the 2- and 3-loops RG prediction for the coupling. At the upper end of the error bar one observes a very close agreement with the non-perturbative coupling. For $\alpha_{\rm{qq}}\lesssim 0.3$ we expect that the perturbative expansion is well behaved and there is no evidence for non-perturbative terms in this region \cite{silvia2}.
The same conclusion is reached considering the relation between $\alpha_{\rm{SF}}(\mu)$ and $\alpha_{\rm{qq}}(\mu)$ at 3 loops
\be\label{qq-sf}
\alpha_{\rm{qq}}(\mu)=h(\alpha_{\rm{SF}}(\mu/s_{0}))+\mathcal{O}([\alpha_{\rm{SF}}(\mu/s_{0})]^{4})
\ee
$$
h(\alpha_{\rm{SF}}(\mu))=\alpha_{\rm{SF}}(\mu)+1.336\,[\alpha_{\rm{SF}}^{3}(\mu)],
$$
where $s_{0}=\lambdasf/\Lambda_{\rm{qq}}$ provides the vanishing of the 2-loop coefficient. As Fig~\ref{alpha_qqsf} shows, at $\alpha_{\rm{qq}}(\mu)\approx 0.2$ the difference $\alpha_{\rm{qq}}(\mu)-h(\alpha_{\rm{SF}}(\mu/s_{0}))$ is not significant; at $\alpha_{\rm{qq}}(\mu)\approx 0.3$ a small difference of order $3\times\alpha^4$ is observable.

\begin{figure}
\includegraphics[width=15pc]{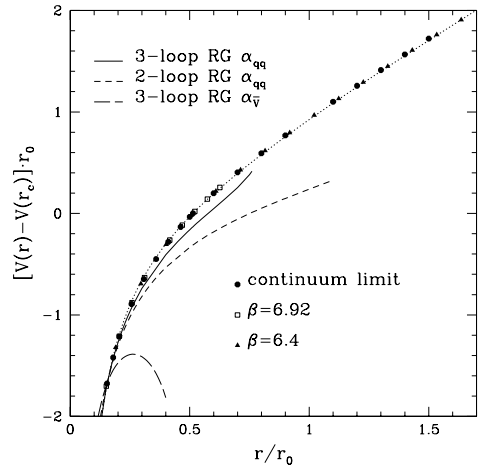}
\vspace{-1cm}
\caption{The potential compared with different perturbative expressions and with the bosonic string model.}\label{potcon}
\end{figure}

Finally, in Fig.~\ref{potcon} the static potential itself is compared with different perturbative approximations. The full line and short dashes correspond to the integration of the force
$$
V(r)=V(0.3\,\rc)+\int_{0.3\,\rc}^{r}dyF(y),
$$
with $F(r)$ from \eq{force} and the 3- and 2-loop RG-solution for $\alpha_{\rm{qq}}$. The long dashes represent \eq{pot_coord} with the 3-loop RG-solution for $\alpha_{\overline{V}}$. As expected, the evaluation of the potential from $\alpha_{\overline{V}}$ is completely unreliable. Furthermore, at large $r$ the potential is expected to be described by a bosonic string model \cite{string}
\be
V(r)=\sigma r+\frac{\pi}{12r},
\ee
which is represented by the dotted line and is in surprising excellent agreement with our results for a long range of $r$.

\section{CONCLUSIONS}
In summary, altough the $\rm V$, $\overline{\rm V}$ and $\rm{qq}$ schemes differ only by kinematics, for the application of perturbation theory they are very different: the RG solution for the coupling suggests that the $\rm{qq}$ scheme provides a reliable perturbative expression up to $\alpha\approx 0.3$. This is confirmed by the lattice results, and there is no evidence for large non-perturbative effects in the short-range potential. To the contrary, perturbation theory works remarkably well where the criterion (i) is satisfied.

\end{document}